\documentclass{article}

\begin{document} 

\begin{titlepage}

\begin{center}
Division Algebras: 26 Dimensions; 3 Families \\
\vspace{.25in}
Geoffrey Dixon \\
gdixon@fas.harvard.edu (until July 1999) \\
gdixon@7stones.com (thereafter) \\
\vspace{.25in}
31 January 1999

\vspace{.25in}
The link of the Division Algebras to 10-dimensional spacetime and one leptoquark family 
is extended to 26-dimensional spacetime and three leptoquark families.
\end{center}

\end{titlepage}

Notation: 
\begin{itemize}

\item \textbf{O} - octonions: nonassociative, noncommutative, basis $\{1=e_{0}, e_{1}, ..., e_{7}\}$;

\item \textbf{Q} - quaternions: associative, noncommutative, basis $\{1=q_{0}, q_{1}, q_{2}, q_{3}\}$; 
\item \textbf{C} - complex numbers: associative, commutative, basis $\{1, i\}$; 

\item \textbf{R} - real numbers. 

\item $\textbf{K}_{L}, \textbf{K}_{R}$ - the adjoint algebras of left and right actions of an 
algebra \textbf{K} on itself.

\item \textbf{K}(2) - 2x2 matrices over the algebra \textbf{K} (to be identified with Clifford algebras);

\item $\mathcal{CL}(p,q)$ - the Clifford algebra of the real spacetime with signature (p+,q-);

\item $^{2}\textbf{K}$ - 2x1 matrices over the algebra \textbf{K} (to be identified with spinor spaces);

\item $\textbf{O}_{L}$ and $\textbf{O}_{R}$ are identical, 
isomorphic to \textbf{R}(8) (8x8 real matrices), \\
64-dimensional bases are of the form $1, \, e_{La}, \, e_{Lab}, \, e_{Labc}$, or 
$1, \, e_{Ra}, \, e_{Rab}, \, e_{Rabc}$, where, for example, if $x \in \textbf{O}$, then 
$e_{Lab}[x] \equiv e_{a}(e_{b}x)$, and $e_{Rab}[x] \equiv (xe_{a})e_{b}$ (see \textbf{[1]});

\item $\textbf{Q}_{L}$ and $\textbf{Q}_{R}$ are distinct, 
both isomorphic to \textbf{Q}, bases \\
$\{1=q_{L0}, q_{L1}, q_{L2}, q_{L3}\}$ and $\{1=q_{R0}, q_{R1}, q_{R2}, q_{R3}\}$;

\item $\textbf{C}_{L}$ and $\textbf{C}_{R}$ are identical, 
both isomorphic to \textbf{C} (so we only need use \textbf{C} itself);

\item \textbf{T} = $\textbf{C}\otimes\textbf{Q}\otimes\textbf{O}$, 64-dimensional;

\item $\textbf{T}_{L}$ = $\textbf{C}_{L}\otimes\textbf{Q}_{L}\otimes\textbf{O}_{L}$, 
isomorphic to $\textbf{C}(16) \simeq \mathcal{CL}(0,9) \simeq \textbf{C} \otimes \mathcal{CL}(0,8)$;

\item \textbf{NOTE}: the only part of $\textbf{T}_{R}$ missing from $\textbf{T}_{L}$ is $\textbf{Q}_{R}$;

\item \textbf{S} = $\textbf{C}\otimes\textbf{Q}\otimes\textbf{Q}\otimes\textbf{Q}
\otimes\textbf{O}\otimes\textbf{O}\otimes\textbf{O}$;

\item $\textbf{S}_{L}$ = $\textbf{C}_{L}\otimes\textbf{Q}_{L}\otimes\textbf{Q}_{L}\otimes\textbf{Q}_{L}
\otimes\textbf{O}_{L}\otimes\textbf{O}_{L}\otimes\textbf{O}_{L}$, 
isomorphic to \\
$\textbf{C}(2^{12}) \simeq \mathcal{CL}(0,25) \simeq \textbf{C} \otimes \mathcal{CL}(0,24)$;

\item \textbf{NOTE}: strictly speaking if we tensor \textbf{Q} and \textbf{O} 3 times each, 
then we should do the same to \textbf{C}, but unlike the former two, 3 tensored copies of 
\textbf{C} can easily be reduced to 1 using projection operators without much evident loss; 
I won't go into this 
now, nor am I completely certain that something might be lost in the simplification of 
$\textbf{T}\otimes\textbf{T}\otimes\textbf{T}$ down to \textbf{S}, but it's worth it 
for the time being.

\item $e_{a}^{(k)}$, a=0,1,...,7, $q_{m}^{(k)}$, m=0,1,2,3, and k=1,2,3, basis elements for the 
three copies of \textbf{O} and \textbf{Q} (similarly for the adjoint algebras);

\end{itemize}
\newpage

Facts (see reference \textbf{[1]}):

\begin{itemize}

\item $\textbf{C}_{L}\otimes\textbf{Q}_{L}(2) \simeq \textbf{C}(4) \simeq 
\textbf{C}\otimes\mathcal{CL}(1,3)$, the Dirac algebra of (1,3)-spacetime 
(the major difference being that the spinor space, $^{2}(\textbf{C}\otimes\textbf{Q})$, 
contains an extra internal SU(2) degree of freedom associated with $\textbf{Q}_{R}$).

\item $\textbf{T}_{L}(2) \simeq \textbf{C}(2^{5}) \simeq \textbf{C}\otimes\mathcal{CL}(1,9)$, 
the Dirac algebra of (1,9)-spacetime (spinor space $^{2}\textbf{T}$; one internal SU(2)).

\item $\textbf{S}_{L}(2) \simeq \textbf{C}(2^{13}) \simeq \textbf{C}\otimes\mathcal{CL}(1,25)$, 
the Dirac algebra of (1,25)-spacetime (spinor space $^{2}\textbf{S}$; 3 internal SU(2)'s).

\end{itemize}

The spinor space of $\textbf{T}_{L}(2)$ can be interpreted as consisting of the direct sum of 
a leptoquark family (2 leptons; 2 quarks; 3 colors) and its antifamily ((1,3)-Dirac spinors; 
see \textbf{[1]}).  
This occurs via a reduction of 
\begin{center} 
$\mathcal{CL}(1,9) \longrightarrow \mathcal{CL}(1,3)\oplus \textbf{(extra bits)}$ 
\end{center} 
using projection operators.  Our goal here 
will be the following: use projection operators to reduce
\begin{equation}
\mbox{(1,25)-spinors} \longrightarrow \mbox{(1,9)-spinors} \longrightarrow \mbox{(1,3)-spinors}, 
\end{equation}
and see what happens to $\mathcal{CL}(1,25)$ under the corresponding algebraic reduction.  
In particular we shall focus on the $\mathcal{CL}(1,25)$ 2-vectors, isomorphic to so(1,25),
and even more particularly we shall focus on the reduction of the transverse subalgebra so(24), 
which reduces as follows:
\begin{equation}
so(24) \longrightarrow so(8)\oplus\textbf{(extra bits)} \longrightarrow 
so(2)\oplus\textbf{(more extra bits)}.
\end{equation}
The extra bits are our penultimate goal.

Define $H = q_{L3}^{(1)}q_{L3}^{(2)}q_{L3}^{(2)}$ and $J = e_{L7}^{(1)}e_{L7}^{(2)}e_{L7}^{(3)}$.  
Then our $\mathcal{CL}(0,24)$ 1-vector basis from $\textbf{S}_{L}$ is the following:
\begin{eqnarray}
Jq_{Lr}^{(1)}q_{L3}^{(2)}; \,\,\,\, Jq_{Lr}^{(2)}q_{L3}^{(3)}; \,\,\,\, Jq_{Lr}^{(3)}q_{L3}^{(1)}; \,\,\,\, r=1,2; \\
ie_{Lp}^{(1)}e_{L7}^{(2)}; \,\,\,\, ie_{Lp}^{(2)}e_{L7}^{(3)}; \,\,\,\, ie_{Lp}^{(3)}e_{L7}^{(1)}; \,\,\,\, p=1,2,3,4,5,6.
\end{eqnarray}
Note that $H$ and $J$ anticommute with all 24 elements listed above.

The corresponding 2-vector basis is
\begin{eqnarray}
q_{L3}^{(k)}, \,\, k=1,2,3; \,\,\,\, q_{Lr}^{(1)}q_{Ls}^{(2)}q_{L3}^{(3)}; 
\,\,\,\, q_{Lr}^{(2)}q_{Ls}^{(3)}q_{L3}^{(1)}; \,\,\,\, q_{Lr}^{(3)}q_{Ls}^{(1)}q_{L3}^{(2)}; \,\, 
r,s=1,2; \\
e_{Lpq}^{(k)}, \,\, k=1,2,3; \,\, e_{Lp}^{(1)}e_{Lq7}^{(2)}e_{Lp}^{(3)}; 
\,\, e_{Lp}^{(2)}e_{Lq7}^{(3)}e_{Lp}^{(1)}; \,\, e_{Lp}^{(3)}e_{Lq7}^{(1)}e_{Lp}^{(2)}; \,\, 
p,q=1,...,6; \\
\mbox{other terms}.
\end{eqnarray}
The elements (5) form a basis for so(6); the elements (6) for so(18); and the elements (5), (6) and (7) 
together for so(24).

\newpage

In order to accomplish the reduction (1), we need some projection operators (recall: we are 
concentrating now on the reduction of the transverse elements, and in particular 
on the reduction of the Lie algebra so(24) given in (5), (6) and (7)).  Define
\begin{eqnarray}
\rho_{\pm}^{(k)} = \frac{1}{2}(1 \pm ie_{7}^{(k)}); \,\,\,\, 
\rho_{L\pm}^{(k)} = \frac{1}{2}(1 \pm ie_{L7}^{(k)}); \,\,\,\, 
\rho_{R\pm}^{(k)} = \frac{1}{2}(1 \pm ie_{R7}^{(k)}); \,\,\,\, k=1,2,3; \\
\lambda_{\pm}^{(k)} = \frac{1}{2}(1 \pm iq_{3}^{(k)}); \,\,\,\, 
\lambda_{L\pm}^{(k)} = \frac{1}{2}(1 \pm iq_{L3}^{(k)}); \,\,\,\, 
\lambda_{R\pm}^{(k)} = \frac{1}{2}(1 \pm iq_{R3}^{(k)}); \,\,\,\, k=1,2,3.
\end{eqnarray}
If $X \in \textbf{S}$, then
\begin{equation}
\rho_{L+}^{(2)}\rho_{R+}^{(2)}\rho_{L+}^{(3)}\rho_{R+}^{(3)}
\lambda_{L+}^{(2)}\lambda_{R+}^{(2)}\lambda_{L+}^{(3)}\lambda_{R+}^{(3)}[X] \equiv 
P[X] = \rho_{+}^{(2)}\rho_{+}^{(3)}\lambda_{+}^{(2)}\lambda_{+}^{(3)}X^{\sim} \equiv pX^{\sim},
\end{equation}
where $X^{\sim} \in \textbf{T}^{(1)} \equiv \textbf{C} \otimes \textbf{Q}^{(1)} \otimes\textbf{O}^{(1)}$, a copy of 
\textbf{T} in \textbf{S}.

The corresponding action on any $U$ in $\textbf{S}_{L}$ is
\begin{equation}
U \longrightarrow PUP
\end{equation}
(note: $P$ is a projection operator, so $P^{2} = P$).  This action reduces the 1-vectors of 
$\mathcal{CL}(0,24)$ to the set
\begin{equation}
pie_{L7}^{(1)}q_{Lr}^{(1)}, \,\, r=1,2; \,\,\,\, pe_{Lp}^{(1)},  \,\, p=1,...,6.
\end{equation}
This is a 1-vector basis for a copy of $\mathcal{CL}(0,8)$, as expected (ie., the transverse 
dimensions of (1,25)-spacetime reduce to the transverse dimensions of (1,9)-spacetime).

The reduction of the 2-vectors (so(24)) is more interesting:
\begin{equation}
so(24) \longrightarrow so(8) \, \times \, (u(1) \times su(3))^{(2)}
\, \times \, (u(1) \times su(3))^{(3)}. 
\end{equation}

The projection operator $\rho_{L+}^{(1)}\rho_{R\pm}^{(1)}$ further reduces the 1-vectors, a reduction 
of (1,9)- to (1,3)-spacetime on the full set.  The reduced Lorentz group, together with a surviving  
su(2) from $\textbf{Q}_{R}^{(1)}$, leaves us with $so(1,25) \longrightarrow so(1,3) \, \times \,$
\begin{equation} 
(u(1) \times su(2) \times su(3))^{(1)} \,
\times \, (u(1) \times su(3))^{(2)} 
\, \times \, (u(1) \times su(3))^{(3)}.
\end{equation}
The spinor reduction can also be done to $\textbf{T}^{(2)}$ and $\textbf{T}^{(3)}$, 
each surviving $u(1) \times su(2) \times su(3)$ associated 
with a different leptoquark family and antifamily.

These ideas are presented as a mathematical exercise.  It has been seen in other theoretical 
arenas that mathematical models of physical reality work out well (resonate with) Lorentz 
spaces of 4, 10 and 26 dimensions.  It is at least interesting that this expansion of a 
model based on 10 dimensions, already shown to have a striking correspondence with the Standard Model 
of one leptoquark family, should continue the correspondence in 26 dimensions.

The preceding mathematical development only scratches the surface.  Continued development would 
shed light on many other matters, including interfamily mixing. 

\newpage
\textbf{References}: \, \\
\vspace{.25in}

[1] G.M. Dixon, \textit{Division Algebras: Octonions, Quaternions, Complex} 

\,\,\, \textit{Numbers, and the 
Algebraic Design of Physics}, (Kluwer, 1994). \\
\vspace{.25in}

[2] G.M. Dixon, www.7stones.com/Homepage/history.html

\end{document}